\documentclass[preprint,amsmath,amssymb]{revtex4-1}
\usepackage{graphicx}
\usepackage{dcolumn}
\usepackage{bm} 
\usepackage{amsmath}
\usepackage{hyperref}
 \usepackage{bigints}   
\usepackage{graphicx}
\usepackage{caption}
\usepackage{subcaption}
\usepackage{setspace}
\usepackage{caption}
\usepackage{color}
\def \O{\mathcal{O}}
\def \o{\omega}
\def \xi{x_{in}}
\def \ti{t_{in}}
\def \xo{x_{out}}
\def \too{t_{out}}
\def \ai{a_{in,\omega}}
\def \adi{a^\dagger_{in,\omega}}
\def \ao{a_{out,\Omega}}
\def \ado{a^\dagger_{out,\Omega}}
\def \x{\vec{x}}
\def \k{\vec{k}}
\def \beq{\begin{equation}}
\def \eeq{\end{equation}}
\begin{document}

\title{Hawking Radiation in Jackiw-Teitelboim Gravity}

\author{Waheed A.Dar} 
\email{waheed.dar1729@gmail.com}
\affiliation{Department of physics \\
National Institute of Technology Srinagar,\\ Kashmir-190006, India}
\author{ Nirmalya Kajuri}%
 \email{nirmalya@iitmandi.ac.in}
\affiliation{%
School of Physical Sciences, Indian Institute of Technology Mandi,\\ Himachal Pradesh, India}
\author{Prince A. Ganai}
\email{princeganai@nitsri.ac.in}
\affiliation{Department of physics \\
National Institute of Technology Srinagar,\\ Kashmir-190006, India}%

\date{\today}

\begin{abstract}
In this paper, we study Hawking radiation in Jackiw-Teitelboim gravity for minimally coupled massless and massive scalar fields. We employ a holography-inspired technique to derive the Bogoliubov coefficients. We consider both black holes in equilibrium and black holes attached to a bath. In the latter case, we compute semiclassical deviations from the thermal spectrum. 
\end{abstract}
\maketitle
\section{Introduction}
There is a long history of studying black holes in two dimensional models in connection to the black hole information loss paradox. The tractability of black hole evaporation in a 2D dilatonic gravity theory makes it a useful toy model to study black hole information loss \cite{Callan:1992rs,Russo:1992ax,deAlwis:1992hv,Giddings:1992ae}. Recently, there has been a surge of renewed interest in a particular 2D dilatonic model-- Jackiw-Teitelboim (JT) gravity \cite{Jackiw:1984je,Teitelboim:1983ux}\footnote{For earlier work on black hole thermodynamics in JT gravity see \cite{Lemos:1993qn,Lemos:1996bq}}. The context of this renewed interest is the ``island solution'' to the black hole information paradox\cite{Almheiri:2019psf,Penington:2019npb,Almheiri:2019yqk}. Here one obtained Page curve for an evaporating black hole in JT gravity, showing that information is not lost(See \cite{Geng:2020qvw,Geng:2020fxl,Geng:2021hlu,Geng:2023zhq} for a critical view of the island proposal). 

Despite this success, it remains unclear how exactly the information escapes from the black hole and appears outside . Clearly, Hawking radiation cannot remain thermal throughout the evaporation and must be modified at late times. It is therefore an important first step to study Hawking radiation in JT gravity. For a conformal field theory, this was done in \cite{Blommaert:2020yeo}. In this paper, we take a first step towards the study of Hawking radiation for minimally coupled scalars in a black hole background in JT gravity. We study this for massive and massless scalars. 

We consider both black holes in equilibrium and out-of-equilibrium. The difference between the two is modeled via boundary conditions. For the former, one employs the usual reflecting boundary conditions, which ensures that the black hole is in equilibrium. To model a black hole out of equilibrium, we follow the setup of \cite{Almheiri:2019psf} and attach a half-Minkowski bath to the black hole. 

To derive Hawking radiation, we follow the usual strategy of computing Bogoliubov coefficients between the modes corresponding to the ``in'' and ``out'' asymptotic regions. The ``in'' asymptotic region is taken to be a Poincare patch while the "out" region corresponds to a black hole. The same strategy was utilized, for example, for CGHS \cite{Giddings:1992ff},BTZ \cite{Hyun:1994na} and Vaidya black holes\cite{Eyheralde:2022ayp}.

As we will see, we will only need to compute the Bogoliubov coefficients up to a factor common to both the $\alpha$ and $\beta$ coefficients  to derive the results that we need, because it is the ratio of the two which is needed to establish thermality. 

For our computations, we will borrow a tool from holography--``boundary representations'' of bulk creation and annihilation operators. Boundary representations were previously used to compute Bogoliubov coefficients in \cite{Dey:2021vke}. 

The key idea of the paper is this: JT gravity can be formulated as a purely boundary theory where the only degree of freedom is the boundary parametrization $f(\tau)$. One can compute $f(\tau)$ for a radiating black hole using the appropriate equations of motion. In the holography-inspired method that we use, it is sufficient to know $f(\tau)$ to compute the Bogoliubov coefficents (up to a common factor, as mentioned above). 

For the case of the black hole in equilibrium, we obtain exact results and find a thermal spectrum, as expected. For the black hole out of equilibrium, we obtain approximate answers in the early and late time regimes. When the black hole settles to a new temperature (the bath temperature), we obtain thermal spectrum in both regimes up to a first approximation. Again, this is the expected result--at early times the black hole is at its original temperature while at late times it settles to a new temperature.

However, we also consider gravitational corrections to the spectrum for early times up to the first order and this shows deviations from thermal spectrum. In the late time regime, we compute corrections away from the strict $t\to \infty$ limit. 

We also consider the case where the bath is at zero temperature i.e the black hole evaporates completely. In this case, we find that there is no Hawking radiation in the late time regime as expected. 

We note that the use of the holographic boundary representations is entirely optional. The main trick one uses here is the simplification of the mode functions in the boundary limit, which was also used in \cite{Belin:2018juv} to compute Bogoliubov coefficients in AdS$_3$.

The paper is structured as follows. The next section provides preliminary information. Here we briefly outline the relevant aspects of JT gravity, and introduce boundary representations. In the third section, we present our results for eternal and evaporating black holes, respectively. We conclude with a summary.

\section{Preliminaries}
\subsection{Jackiw-Teitelboim Gravity}
Jackiw-Teitelboim gravity(JT gravity) is a 2D dilatonic model of gravity. The action of JT gravity minimally coupled to matter is given by:
\begin{eqnarray}
  I[g,\Phi]=I_{\text{grav}}[g, \phi]  + I_{\text{matter}}
\end{eqnarray}
Here $\phi$ denotes the dilatonic field. 
The gravitational action is given by:
\begin{align} \notag
I_{\text{grav}}[g, \phi]=\frac{-1}{16 \pi G_{N}} \int_{\mathcal{M}} \sqrt{g} & \Phi \left(  R + 2 \right) - \frac{1}{8 \pi G_{N}} \int_{\partial \mathcal{M}} \sqrt{h}  \Phi (K - 1) 
\\ &-\frac{\phi_0}{16 G_N \pi }\left(\int_{\mathcal{M}} \sqrt{g}R+2 \int_{\partial \mathcal{M}} \sqrt{h} K\right)
\end{align}

JT gravity has been widely studied because it provides a concrete realization of the NAdS$_2$/CFT correspondence. We refer the reader to \cite{Mertens:2022irh} for a recent review. 

The dilaton equation of motion imposes the two-dimensional geometry to be locally AdS$_{2}$. The geometry is thus non-dynamical. 

The only dynamical gravitational degree of freedom can be taken to be the freedom to reparametrize the boundary time $\tau \to f(\tau)$, up to $SL(2,R)$ transformations. Parametrizations related by $SL(2,R)$ transformations:
\beq 
f(\tau)\to \frac{af(\tau)+b}{cf(\tau)+d}
\eeq
are physically equivalent. 

JT gravity can then be formulated as a purely boundary theory given by the Schwarzian action\cite{Maldacena:2016upp,Engelsoy:2016xyb}:
\begin{eqnarray}
I_{\text{grav}} = \frac{\Phi_{r}}{8 \pi G_{N}} \int d\tau \{f(\tau), \tau\}
\end{eqnarray}

where the Schwarzian is defined as usual:
$$
\{f(\tau), \tau\}=\frac{f^{\prime \prime \prime}}{f^{\prime}}-\frac{3}{2}\left(\frac{f^{\prime \prime}}{f^{\prime}}\right)^2
$$
The effect of backreaction on the gravitational sector from interaction with matter is to modify the parametrization $f(\tau)$. This can be determined from the equation of motion of JT gravity coupled to matter: 
\begin{eqnarray}
    \partial_{\tau}E=T_{x^{-}{x^{-}}}-T_{x^{+}{x^{+}}} \label{eq:energy}
\end{eqnarray}
where $T_{x^{-}{x^{-}}}$ and $T_{x^{+}{x^{+}}}$ are the ``in'' and ``out'' fluxes of matter sector and $E(\tau)$ is the ADM  energy of bulk spacetime given by:
\begin{eqnarray}
E(\tau) = \frac{-\Phi_{r}}{8 \pi G_{N}} \{f(\tau), \tau\}
\label{eq:dilaton}
\end{eqnarray}

Here, $\phi_r$ is the value of the dilaton at the boundary.

\subsection{Boundary Representations}

According to AdS/CFT dictionary, every bulk field has a boundary dual. For a free scalar $\phi(r,\x)$ of mass $m$ in the bulk, the corresponding boundary operator is a scalar primary $\O(\x)$ of conformal dimension $\Delta =d/2 + \sqrt{d^2/4+m^2}$. The two are related through the extrapolate dictionary\cite{Banks:1998dd,Harlow:2011ke}:

\begin{align}\label{xtra}
\lim_{r\to \infty}r^{n\Delta} &\langle\phi(r,\x_1 )\phi(r,\x_2 )...\phi(r,\x_n )\rangle = \langle 0|\O(\x_1 )\O(\x_2 )..\O(\x_n )|0\rangle 
\end{align} 

From \eqref{xtra} it can be shown\cite{Dobrev:1998md, Bena:1999jv,Hamilton:2005ju,Papadodimas:2012aq} that the creation and annihilation operators of the field $\phi$ satisfies the following:
\begin{align}\notag
a_{\k} &= \O_{\k}/c_{\k}  \\ \label{anni}
a^\dagger_{\k} &= \O^\dagger_{\k}/c_{\k}  
\end{align} 
where 
\begin{align}\notag
  \O_{\k} &\equiv \int d^dx \, \O(\x)e^{-i\k\cdot\x} \\
  \O^\dagger_{\k} &\equiv \int d^dx \, \O(\x)e^{i\k\cdot\x}
\end{align}

and \beq 
c_{\k} =\lim_{r\to \infty}r^{\Delta} f_{k}(r).
\eeq 
 where $f_{\k}$ are the normalized mode functions. We will not need to compute $c_{\k}$ explicitly. The key point is that the same factor $c_{\k}$ appears in both the annihilation and creation operators. 

The equality relation in \eqref{anni} holds within correlation functions, as in \eqref{xtra}. The operators $\frac{\O_{\k}}{c_{\k}}, \frac{\O^\dagger_{\k}}{c_{\k}} $ are called ``boundary representations'' of bulk creation and annihilation operators $a_{\k} ,a^\dagger_{\k}$.

We refer the reader to \cite{Kajuri:2020vxf} for a review of boundary representations. 
\section{Computation of Bogoliubov coefficients}

Our aim is to compute the Bogoliubov coefficients between the modes corresponding to ``in" and ``out" asymptotic regions in a 2D black hole. The coordinates are denoted as $(\ti,\xi)$ and $(\too,\xo)$ respectively.

The corresponding mode expansions are written schematically as:
\begin{align} 
\phi&=\int {d \omega} \left[f^{in}_\o\ai + \left(f^{in}_\omega\right)^* \adi\right]\\
&=\int  {d \Omega} \left[f^{out}_\Omega \ao + \left(f^{out}_\Omega \right)^* \ado\right]
\end{align} 

The out modes can be written as linear combinations of in modes as:
\begin{align}
f^{out}_\Omega= \sum_{\o} \alpha_{\o \Omega}f^{in}_\o +\beta_{\o \Omega}\left(f^{in}_\o\right)^*
\end{align}
Here \(\alpha_{\omega \Omega}\) and \(\beta_{\omega \Omega}\) are the Bogoliubov coefficients.

When the `in' vacuum is populated in terms of `out' particles, we have particle creation. The condition for particle creation is:
\beq 
|\beta_{\o\Omega}|^2>0.
\eeq 
The condition for thermal spectrum is that Bogoliubov coefficients satisfy the property:
\beq \label{thermal}
|\alpha_{\o \Omega}| = e^{\beta \Omega}|\beta_{\o \Omega}|.
\eeq
We then have the thermal spectrum:
\beq 
_{in}\langle 0| N_\Omega|0\rangle_{in}= \frac{e^{-\beta \Omega}}{1-e^{-\beta \Omega}}
\eeq
Here $N_\Omega$ is the number operator.
We can also write the Bogoliubov coefficients in terms of the creation and annihilation operators:
\begin{align} \label{aai}
\ai &= \sum_\Omega \alpha_{\o \Omega} \ao +\beta^*_{\o \Omega}\ado\\
\adi &= \sum_\Omega  \alpha^*_{\o \Omega} \ao +\beta_{\o \Omega}\ado
\end{align} 
At this stage we can employ the boundary representations of creation and annihilation operators introduced earlier: 
\begin{align} \label{ai}
\ai &= \O^{in}_\o/c_\o,\,\adi=\O^{in}_{-\o}/c_\o \\
\ao &= \O^{out}_\o/c_\o,\,\ado=\O^{out}_{-\o}/c_\o
\end{align}
where 
\begin{align}
  \O^{in}_\o= \int d\ti e^{i\o \ti} \O(\ti)\\
  \O^{out}_\o= \int d\too e^{-i\o \too} \O(\too)
\end{align}
Similarly for the $\O_{\o}^\dagger$ operators. 

The boundary operators $\O^{in}_\o,\O^{out}_\o$ can be related as follows:
\begin{align}
\O^{in}_\o&=\int d\ti e^{i\o \ti} \O(\ti)=\int d\too|d\ti/d\too|^{1-\Delta}\O(\too)e^{i\o \ti}
\end{align}
where we have used $\O(\ti)=|d\ti/d\too|^{-\Delta}\O(\too)$.
By Fourier expanding $|d\ti/d\too|^{1-\Delta}e^{i\o \ti(\too)}$ in the modes $e^{i\Omega \too}$ we obtain:
\beq 
\O^{in}_\o= \sum^{\infty}_{\Omega=0} \left(C(\o,\Omega)\O^{out}_\Omega + C(\o,-\Omega)\O^{out}_{-\Omega}\right)
 \eeq
where
\beq 
C(\o,\Omega)=\int d\too|d\ti/d\too|^{1-\Delta}e^{i\o \ti}e^{-i\Omega \too}
\eeq 
Note that the only input needed to compute $C(\o,\pm\Omega)$ is the functional dependence of $t_{in}$ on $t_{out}$.
We have then from \eqref{aai} and \eqref{ai} that
\begin{align}
\alpha_{\o \Omega}&=C(\o,\Omega)/c_\o\\
\beta^*_{\o \Omega}&=C(\o,-\Omega)/c_\o
\end{align}

When the bath is attached, the logic is slightly different: The logic of our derivation is as follows. When the bath is attached, the bulk field is given by 

 \begin{equation}
   \phi(r,\vec{x}) = A\phi_N(r,\vec{x}) + B\phi_{NN}(r,\vec{x}) 
 \end{equation}

 where the subscripts denote the normalizable and non-normalizable modes. These have different fall-offs: 

$$\lim_{r \to \infty} \phi_N(r,\vec{x}) \to r^{-\Delta_{+} }\mathcal{O}_{\Delta_{+} }(\vec{x}).$$

$$\lim_{r \to 0} \phi_{NN}(r,\vec{x}) \to r^{-\Delta_{-}} \mathcal{O}_{\Delta_{-} }(\vec{x}). $$
where 
$$\Delta_\pm =d/2 \pm \sqrt{d^2/4+m^2}$$
It follows that one can still use the dominant fall-off and take the $r\to \infty$ limit to get:

$$ \lim_{r \to \infty}r^{-\Delta_{+}} \phi_N(r,\vec{x}) \to A\mathcal{O}_{\Delta_{+} }(\vec{x}).$$

We use this relation to obtain the Boguliubov coefficients as described in the paper for the case of the bath. The difference between the bath and the no-bath case is reflected in the fact that the coefficient $A$ remains, and this is why we are able to compute only the ratio of the Bogliubov coefficients in this case

We should note that the AdS/CFT language is not strictly necessary for our derivation. We are simply using the fact that the bulk mode functions simplify in the boundary limit ($f_\omega(r, t)\sim r^{-\Delta} e^{i \omega t}$ for either mode). The derivation of Bogoliubov coefficients is morally the same as the derivation presented in \cite{Belin:2018juv} . 

We now proceed to compute Bogoliubov coefficients for specific cases.

\subsection{Black Holes in Equilibrium}
First we consider an eternal black hole at equilibrium. This is achieved by implementing reflecting boundary conditions at the boundary. 
The ``in'' and ``out'' asymptotic regions correspond to Poincare and black hole (or AdS-Rindler) metrics respectively.
The Poincare metric is given by:
\beq 
d s^2=1 / Z^2\left(-d T^2+d Z^2\right)
\eeq
and the AdS-Rindler metric is given by:
\beq 
ds^2 = \frac{4\pi^2}{\beta^2} \frac{(-dT^2 + d\rho^2)}{\sinh^2 \left( \frac{2\pi}{\beta} \rho \right)} 
\eeq 
The transformation relating these two is given by,
\beq \label{coord}
T=\frac{\beta}{2\pi} e^{\frac{2\pi}{\beta } \tau} \cosh(\rho),\quad Z=\frac{\beta}{2\pi} e^{\frac{2\pi}{\beta }\tau} \sinh(\rho)
\eeq
At the boundary $Z=0 \leftrightarrow$ $\rho=0$. The two time coordinates are related by:
$$
T=\frac{\beta}{2\pi}  e^{\frac{2\pi}{\beta}\tau}
$$
The boundary dual to the bulk field will be a conformal primary $\O$. We define the following boundary operators for the Poincare patch: 
 \begin{align}
 \O^{P}_{\omega} &= \int_{-\infty}^{+\infty} dT \, \O(T) e^{i \omega T} \\
 \O^{P}_{-\omega} &= \int_{-\infty}^{+\infty} dT \, \O(T) e^{-i \omega T} \\
 \end{align}
While for AdS-Rindler:
\begin{align} 
 \O_{\Omega}^{\text{Rind}} &= \int_{-\infty}^{+\infty} d\tau \, \O(\tau) e^{i \Omega \tau} \\
 \O_{-\Omega}^{\text{Rind}} &= \int_{-\infty}^{+\infty} d\tau \, \O(\tau) e^{-i \Omega \tau}
\end{align}
The Poincare modes are the `in' modes, and the AdS-Rindler modes are the `out' modes. 

Just as Minkowski modes can be written as a sum over Rindler modes on two Rindler wedges (left and right), Poincare modes too will divide into a sum over two modes in two AdS-Rindler wedges: 
\begin{align} \notag
\O^{P}_{\omega} &= \int_{-\infty}^{+\infty} dT \, \O(T) e^{i \omega T} = \int_{-\infty}^{0} dT \, \O(T) e^{i \omega T}+\int_{0}^{+\infty} dT \, \O(T) e^{i \omega T}\\
&=\int_{-\infty}^{+\infty} d\Omega\, C_L(\omega, \Omega) \O^{Rind, L}_\Omega  +\int_{-\infty}^{+\infty} d\Omega\, C_R(\omega, \Omega) \O^{Rind,R}_\Omega   
\end{align} 
where we have labelled the two wedges $R,L$. The exterior of the black hole lies in a single wedge however, and we need only consider one wedge. We therefore only consider the "right" wedge and drop the label $R$.

We need to compute:
\begin{equation}
C(\omega, \Omega)=\int_{-\infty}^{\infty} d \tau \left| \frac{dT}{d\tau}\right|^{1-\Delta} e^{i \omega T(\tau)} e^{-i \Omega \tau}
\end{equation}
First, we consider the case of the massless field. In this case $\Delta=1$. 

We have: 
\begin{eqnarray*}
C(\omega, \Omega) &=& \int_{-\infty}^{\infty} d\tau \, e^{i\frac{\omega \beta}{2\pi} e^{\frac{2\pi}{\beta}\tau}} e^{-i \Omega \tau}
\end{eqnarray*}
Performing the integration, we obtain:
\begin{eqnarray*}
C(\omega, \Omega) =\frac{\beta}{2\pi} {e^{i \Omega \frac{\beta}{2\pi}  \ln (-i\frac{\omega \beta}{2\pi})}} \Gamma(-i \frac{\Omega \beta}{2\pi})
\end{eqnarray*}
We have, therefore:
\begin{align}
c_\o \alpha_{\omega \Omega} &= \frac{\beta}{2\pi} e^{i \Omega \frac{\beta}{2\pi} \ln \left(-i\frac{\omega \beta}{2\pi}\right)} \Gamma\left(-i \frac{\Omega \beta}{2\pi}\right)= \frac{\beta}{2\pi} e^{\frac{\Omega \beta }{2}} e^{i \Omega \frac{\beta}{2\pi} \ln \left(\frac{\omega \beta}{2\pi}\right)}\Gamma\left(-i \frac{\Omega \beta}{2\pi}\right) \\
c_\o \beta_{\omega \Omega}^{*} &= \frac{\beta}{2\pi} e^{-i \Omega \frac{\beta}{2\pi} \ln \left(-i\frac{\omega \beta}{2\pi}\right)} \Gamma\left(i \frac{\Omega \beta}{2\pi}\right)= \frac{\beta}{2\pi} e^{-\frac{\Omega \beta }{2}} e^{i \Omega \frac{\beta}{2\pi} \ln \left(\frac{\omega \beta}{2\pi}\right)}\Gamma\left(i \frac{\Omega \beta}{2\pi}\right)
\end{align}
Using $\Gamma(\Bar{z})=\Bar{\Gamma}(z)$, we get:
$$
\alpha_{\omega \Omega} = e^{\frac{\beta}{2} \Omega}
 \beta_{\omega \Omega} 
$$

which satisfies the thermality condition \eqref{thermal}, as expected. We note that we did not need to compute the constant $c_\o$ to deduce thermality as it is common to both the coefficients. The Bogoliubov coefficients for this case where previously computed in \cite{Cadoni:1994uf}. Our results match theirs up to a phase.  

We now consider the case of a massive scalar. In this case $\Delta \neq 1$ and we have: 
\beq
C(\omega,\Omega) = \int_{-\infty}^{\infty} d\tau \left(e^{\frac{2\pi \tau}{\beta}}\right)^{1-\Delta} e^{i \frac{\omega \beta}{2\pi} e^{\frac{2\pi \tau}{\beta}}} e^{-i \Omega \tau}
\eeq
This integrates to give:
$$
C({\omega, \Omega})=e^{(1-\Delta-i \Omega \beta/ 2\pi)  \ln (i \omega \beta/ 2\pi)} \Gamma(1-\Delta-i \Omega \beta / 2\pi)
$$
Then up to the $c_\o$ factor, we have:
\begin{align}
\alpha_{\omega \Omega} &\sim e^{(\Omega \beta/ 4)}e^{i(1-\Delta)\frac{\pi}{2}}  
e^{(1-\Delta-i \Omega \beta/ 2\pi)\ln ( \omega \beta/ 2\pi)} \Gamma(1-\Delta-i \Omega \beta / 2\pi)\\
\beta^*_{\omega \Omega} &\sim e^{(-\Omega \beta/ 4)}e^{-i(1-\Delta)\frac{\pi}{2}}
e^{(1-\Delta+i \Omega \beta/ 2\pi) \ln ( \omega \beta/ 2\pi)} \Gamma(1-\Delta+i \Omega \beta / 2\pi)
\end{align}
Thus the Bogoliubov coefficients for the massive case also satisfy:
\beq \label{cond}
\alpha_{\omega \Omega}  =e^{\frac{\beta}{2} \Omega}
 \beta_{\omega \Omega} 
\eeq 
The radiation is thermal, again as expected. 

\subsection{Black Hole Attached to bath}

We now consider the case of a black hole that is not in equilibrium. The way to achieve this is to start with a black hole in equilibrium and then attach a half-Minkowski ``bath''  to it at temperature $\tilde{\beta}$, which exchanges energy with the black hole\cite{Almheiri:2019psf}. The black hole eventually settles at the new temperature $\tilde{\beta}$. If the bath is at zero-temperature, the black hole evaporates completely.

As we have discussed earlier, JT gravity can be formulated with the boundary time as the sole dynamic variable. The effect of attaching the bath and the subsequent escape of radiation is a reparametrization of the boundary time $\tau\to f(\tau)$. The reparametrization $f(\tau)$ determines the ADM energy of the black hole and can be determined through the equations of motion.From equation~\ref{eq:energy}
\begin{eqnarray}
    \partial_{\tau}E=T_{y^{+}{y^{+}}}-T_{y^{-}{y^{-}}} 
\end{eqnarray}
The RHS of the above equation determines the net flux of energy at the boundary . \( T_{y^{-}y^{-}} \) is the outgoing flux, i.e., Hawking radiation, and \( T_{y^{+}y^{+}} \) is the incoming flux at the boundary, provided by the bath. In order to transform the stress-energy tensor components from Poincaré coordinates \(x^{\pm}\) to black hole (AdS-Rindler) coordinates \( y^{\pm} \), we use the stress-energy tensor anomaly from 2D CFT, given as:
\begin{eqnarray}
   T_{x^{\pm}x^{\pm}} = \left( \frac{\partial y^{\pm}}{\partial x^{\pm}} \right)^2 
   \left[ T_{y^{\pm}y^{\pm}} + \frac{C}{24\pi} \{ x^{\pm}, y^{\pm} \} \right]
\label{eq:anomaly}
\end{eqnarray}
where C is the central charge of CFT. The map \( x^{\pm} = f(y^{\pm}) \) is our new parametrization of the boundary for transparent boundary conditions, which we need to find .
The stress-energy tensor, for incoming modes, is 
\begin{eqnarray}
  E_{\tilde{\beta}} = T_{y^+ y^+} = \frac{\pi C}{12 \tilde{\beta}^2}
\end{eqnarray}

while for outgoing modes, it is 
\begin{eqnarray}
    T_{y^- y^-} = \frac{-C}{24\pi} \{f(y^-), y^-\}
\end{eqnarray}

then from \ref{eq:energy} and \ref{eq:anomaly} , we get
\begin{eqnarray}
    \partial_{\tau} E = \frac{\pi C}{12 \tilde{\beta}^2}+ \frac{C}{24\pi} \{f(\tau), {\tau}\}
    \label{tilde}
\end{eqnarray}
The above equation is solved with boundary condition \(E_{0}=E_{\beta}\) , here \(E_{\beta}\) is the black hole energy.
Using above equation \ref{tilde} and energy in terms of schwarzian\ref{eq:dilaton},we get a differential equation:
\begin{eqnarray}
    -\frac{\Phi_{r}}{8 \pi G_{N}} \partial_{\tau}\{ f(\tau), \tau \} =  \frac{\pi C}{12 \tilde{\beta}^2}+\frac{C}{24\pi} \{ f(\tau), \tau \} 
    \implies \{ f(\tau), \tau \} \propto e^{-k\tau}
\end{eqnarray}
The solution of this differential equation is given by \cite{Almheiri:2019psf,Hollowood:2020cou}:
\beq \label{sol}
f(\tau)= \frac{e^{\frac{4\pi}{\beta k}}}{\pi}\frac{K_{\nu}(z)}{I_{\nu}(z)}
\eeq
where we have defined 
\begin{eqnarray}
 z=\frac{2\pi}{\tilde{\beta}k} \sqrt{\frac{\tilde{\beta}^2}{\beta^2} - 1} \,~ e^{-k{\tau}/2} 
\end{eqnarray}

and we take \(\tilde{\beta} \gg \beta\).
then we have ,
 \[k=\frac{G_N C}{3 \Phi_r},\quad z= \frac{2\pi}{\beta k} e^{-k\tau/2} , ~\nu=\frac{2\pi}{\tilde {\beta} k}\] $K_{\nu}(z),I_{\nu}(z)$ are Bessel functions.

But as we saw in the last section, $f(\tau)$ is all we need to know to determine the Bogoliubov coefficients (up to an overall constant). We will use this technique to carry out the computations at the early and late regimes. 

Before we proceed, it is important to clarify the meaning behind our computations at these regimes. The `out' vacuum is not a fixed state in the evaporating case, it keeps changing with time. The Bogoliubov coefficients we find are between the approximate vacuum states at early and late time regimes. This is why, even though the approximate $f(\tau)$ corresponding to these vacua are valid for small time windows, we can take the integral to cover the entire time slice. 

We now proceed with the following computations.

\textbf{Early Time Regime:}
We take this to be defined as $\tau \ll k^{-1}|\log{{\beta}k}|$. In this regime $z\to \infty$. 
We can therefore utilize the asymptotic limit of Bessel functions with $\nu$ fixed\cite{Abramowitz}:
\begin{align}
\lim_{z \to \infty} I_{\nu}(z) \approx \frac{e^{z}}{(2\pi z)^\frac{1}{2}},\quad
\lim_{z \to \infty} K_{\nu}(z) &\approx \frac{\pi^{\frac{1}{2}}e^{-z}}{(2 z)^\frac{1}{2}}
\end{align}
which gives
\beq \label{ratio1}
\frac{K_{\nu}(z)}{ I_{\nu}(z)} \approx \pi e^{-2z}
\eeq
Note that $\nu$-dependence has dropped out in this limit.
We have then:
\beq \label{zsim}
f(\tau)=e^{\frac{4\pi}{\beta k}}e^{- \frac{4\pi}{\beta k} e^{-k\tau/2}  }
\eeq 
We now use a second approximation where we take the effective gravity-matter coupling $k \ll 1$. In other words, we restrict to the semiclassical regime \cite{Hollowood:2020cou}. We use this to expand $e^{f(\tau)}$--the quantity that enters the computation of Bogoliubov coefficients--order by order in $k$. 
Using \eqref{zsim}, we have:
\beq \label{par}
e^{i \omega f(\tau)}= e^{i \omega e^{\frac{2 \pi \tau}{\beta}}}+k \left(i \o \frac{\pi \tau^2}{\beta}e^{\frac{2 \pi \tau}{\beta}}e^{i \o e^{\frac{2 \pi \tau}{\beta}}}\right)+\mathcal{O}(k^2)
\eeq
Up to the zeroeth order in $k$, we have then:
\beq 
f(\tau)= e^{\frac{2 \pi \tau}{\beta}}
\eeq
This is the same parametrization for the boundary we had in reflecting boundary conditions(up to a constant), therefore we will get the same Bogoliubov coeficients in both the massless and massive case. The thermal spectrum we obtain is expected because we are at an early stage of the evaporation, and we have ignored gravitational corrections to the spectrum. 

We can compute the corrections to the thermal spectrum order-by-order in $k$. Here we only compute the first order correction for the massless case, but it should be possible, if not particularly insightful, to compute them to arbitrarily high orders. 

The first order correction in $k$ to $C(\omega,\Omega)$ is given by the integral:
\beq 
 C^{(1)}(\omega,\Omega)=\frac{i \pi \o k }{\beta}\int_{-\infty}^{\infty} d\tau \,\tau^2e^{\frac{2 \pi \tau}{\beta}}e^{i \o e^{\frac{2 \pi \tau}{\beta}}} e^{-i \Omega \tau}
\eeq
By substituting $e^\tau=z$, we can convert this integral into a known Mellin transform\cite{Bateman} and obtain: 
\begin{align*} 
&C^{(1)}(\omega,\Omega)=\\
&\frac{i \omega k \beta^2}{8\pi^2}
 (i \omega )^{-1+\frac{i \Omega \beta }{2 \pi }}\left( \log ^2(-i \omega ) \Gamma \left(1-\frac{i \Omega \beta }{2 \pi }\right)+(-i \omega )^{-1+\frac{i \Omega \beta }{2 \pi }} \left(\Gamma \left(1-\frac{i \Omega \beta }{2 \pi }\right) \psi ^{(0)}\left(1-\frac{i \Omega \beta }{2 \pi }\right)^2 \right. \right.\\&+ \left. \left. \Gamma \left(1-\frac{i \Omega \beta }{2 \pi }\right) \psi ^{(1)}\left(1-\frac{i \Omega \beta }{2 \pi }\right)\right)-2 (-i \omega )^{-1+\frac{i \Omega \beta }{2 \pi }} \log (-i \omega ) \Gamma \left(1-\frac{i \Omega \beta }{2 \pi }\right) \psi ^{(0)}\left(1-\frac{i \Omega \beta }{2 \pi }\right)\right)
\end{align*}
where $\psi$ is the Polygamma function. 

As can be checked, the condition \eqref{thermal} for thermal spectrum is no longer satisfied once the corrections are incorporated.

\textbf{Late time regime:} We take the late time regime to be $\tau \gg k^{-1}|\log{\beta k}|$. In this regime $ z \rightarrow 0$. We will not use the $k\ll 1$ approximation here, hence our results will not be restricted to the semiclassical regime.

There are two separate cases to consider. First we consider the case where the bath is at non-zero temperature i.e $\nu \neq 0$, and black hole settles to the bath temperature of $\tilde{\beta}$

We have in the late time limit, for a given $\nu$\cite{Abramowitz}:
\[\lim_{z \to 0} K_{\nu}(z) \approx \frac{\Gamma{(\nu)}}{2(\frac{z}{2})^{\nu}}\] and 
\[\lim_{z \to 0} I_{\nu}(z) \approx \left(\frac{z}{2}\right)^{\nu} \frac{1}{\Gamma(\nu +1)}\] 

then we have 
\begin{align} 
f(\tau)= \frac{e^{\frac{4\pi}{\beta k}}}{\pi}\frac{K_{\nu}(z)}{I_{\nu}(z)}\approx \frac{e^{\frac{4\pi}{\beta k}}}{\pi} \nu ~[\Gamma(\nu)]^2~ 2^{2\nu-1}~{z^{-2\nu}}= \frac{e^{\frac{4\pi}{\beta k}}}{\pi}  \left({\frac{\beta k}{2\pi}}\right)^{2\nu} \nu{[\Gamma(\nu)]^2~ 2^{2\nu-1}}~e^{{\frac{2\pi}{\tilde{\beta}}}\tau}
\end{align}

Introducing the constant
\[
C_1 =\frac{e^{\frac{4\pi}{\beta k}}}{\pi} \left({\frac{\beta k}{2\pi}}\right)^{2\nu} \nu{[\Gamma(\nu)]^2~ 2^{2\nu-1}}
\]

We write the parametrization as:
\begin{align}
f(\tau) = C_1 ~e^{\frac{2\pi}{\tilde{\beta}} \tau}
\end{align}

The parametrization is same as \eqref{par} up to a constant, so we get similar results once again. 

For the massless case, we have 

\begin{align}
c_\o \alpha_{\omega \Omega} &=  \frac{\tilde{\beta}}{2 \pi}e^{\frac{\Omega \tilde{\beta}}{2 \pi} }e^{i \frac{\Omega \tilde{\beta}}{2 \pi} \ln( \omega C_1) } \Gamma(-\frac{i \Omega \tilde{\beta}}{2 \pi})\\
c_\o \beta_{\omega \Omega} &=  \frac{\tilde{\beta}}{2 \pi}e^{-\frac{\Omega \tilde{\beta}}{2 \pi} }e^{i \frac{\Omega \tilde{\beta}}{2 \pi} \ln( \omega C_1) } \Gamma(-\frac{i \Omega \tilde{\beta}}{2 \pi})
\end{align}

and for the massive case:
\begin{align}
c_\o \alpha_{\omega \Omega} &= e^{\left( \frac{\Omega \tilde{\beta}}{2\pi}\right)} \left( \frac{\tilde{\beta}}{2\pi} \right)^{\Delta} C_1^{1-\Delta}e^{i \pi \Delta} e^{\left( \frac{i \Omega\tilde{\beta}}{2\pi} + \Delta \right) \ln(i \omega C_1)} \Gamma\left( \frac{-i \Omega \tilde{\beta}}{2 \pi} - \Delta + 1 \right)\\
c_\o \beta_{\omega \Omega} &= e^{\left( -\frac{\Omega \tilde{\beta}}{2\pi}\right)} \left( \frac{\tilde{\beta}}{2\pi} \right)^{\Delta} C_1^{1-\Delta}e^{i \pi \Delta} e^{\left( \frac{i \Omega \tilde{\beta}}{2\pi} + \Delta \right) \ln(i \omega C_1)} \Gamma\left( \frac{-i \Omega \tilde{\beta}}{2 \pi} - \Delta + 1 \right)
\end{align}

Both these cases satisfy \eqref{cond} as before, and therefore have thermal spectra. This too is expected, since at late times the black hole settles down to a temperature $\tilde{\beta}$. 

We can go beyond the strict late time limit by considering the asymptotic expansions of the Bessel functions for large $z$: 
\begin{align*}
 K_\nu(z) \approx  z^{-\nu } \left(2^{\nu -1} \Gamma (\nu )-\frac{2^{\nu -3} \Gamma (\nu ) z^2}{\nu -1}+\frac{2^{\nu -6} \Gamma (\nu ) z^4}{(\nu -2) (\nu -1)}+\dots\right) 
\end{align*}

\begin{align*}
I_{\nu}(z) \approx z^{\nu } \left( \frac{2^{-\nu }}{\Gamma (\nu + 1)} + \frac{2^{-\nu - 2} z^2}{(\nu + 1) \Gamma (\nu + 1)} + \frac{2^{-\nu - 5} z^4}{(\nu + 1)(\nu + 2) \Gamma (\nu + 1)} + \dots \right)
\end{align*}

We have then: 
\begin{align}
 f(\tau) \approx z^{-2 \nu}( C_1+ C_2 z^2 +C_4 z^4+ \dots) 
\end{align}
where $C_2,C_4$ are constants depending on $\nu$. 

We display $C_2$ explicitly since we will use it: 
\begin{align}
C_2&=-\frac{e^{\frac{4 \pi}{\beta k}}}{\pi}\frac{2^{2 \nu -2} \nu   \Gamma (\nu ) \Gamma (\nu +1)}{(\nu -1) (\nu +1)}
\end{align}

Then expanding $e^{i \o f(\tau)}$ around $z=0$ we get: 
\beq 
e^{i \omega f(\tau)}=e^{i \o C_1e^{\frac{2\pi \tau}{\tilde{\beta}}}}+ C_1C_2  i\o  e^{\frac{2\pi \tau}{\tilde{\beta}}} e^{i \o C_1 e^{\frac{2\pi \tau}{\tilde{\beta}}}}\frac{4\pi}{\beta^2 k^2}e^{-k \tau}+\dots
\eeq 

It is straightforward to compute the corrections to Bogoliubov coefficients to arbitrary orders. Here we compute the first order correction for the massless case: 
\beq 
C^{(1)}(\omega,\Omega) =  i\o C_1 C_2\frac{\tilde{\beta}}{2 \pi}e^{\frac{\Omega \tilde{\beta}}{2 \pi} }e^{i \frac{\Omega \tilde{\beta}}{2 \pi} \ln( \omega C_1 ) } \Gamma\left(1-\frac{(k+i \Omega) \tilde{\beta}}{2 \pi}\right)
\eeq

As can be easily checked, this is still thermal at temperature $1/\tilde{\beta}$. Thus there is no deviation from thermal spectrum in this regime. 

Note that this is  different from the semiclassical correction we considered for the early time regime. There we considered an expansion in $k$, here we consider an expansion in $z$. That is, we extend back in time from the exact late time limit.

We will now treat the case where the black hole evaporates completely. This happens when the black hole is coupled to a zero temperature bath, i.e $\nu=0$. In this case, the Bessel functions have the following limiting behavior\cite{Abramowitz}:
\begin{align}
\lim _{z \rightarrow 0} K_0(z) &\approx-\ln z\\
I_0(z)& \approx 1
\end{align}
The parametrization is: 
$$
f(\tau)=\frac{e^{\frac{4 \pi}{\beta k}}}{\pi} \frac{K_0(z)}{I_0(z)} \approx  \frac{e^{\frac{4 \pi}{\beta k}}}{\pi}\left[\log \left(\frac{\beta k}{2 \pi}\right)+\frac{k \tau}{2}\right]
$$

For the massless case we get:
\begin{align}
C(\omega, \Omega) &= e^{\frac{i\omega}{\pi} \left[ \log \left( \frac{\beta k}{2 \pi} \right) + e^{\frac{4 \pi}{\beta k}} \right]} \delta \left( \frac{\omega k}{2 \pi} - \Omega \right)\\
C(\omega, -\Omega)&= e^{\frac{i\omega}{\pi} \left[ \log \left( \frac{\beta k}{2 \pi} \right) + e^{\frac{4 \pi}{\beta k}} \right]} \delta \left( \frac{\omega k}{2 \pi} - \Omega \right)\label{c1}
\end{align}

Then we have 
\begin{align}
 c_\o \alpha_{\omega \Omega} &= e^{\frac{i\omega}{\pi} \left[ \log \left( \frac{\beta k}{2 \pi} \right) + e^{\frac{4 \pi}{\beta k}} \right]} \delta \left( \frac{\omega k}{2 \pi} - \Omega \right) \\
 \beta_{\omega \Omega}&=0 \label{zero}
\end{align}
Here \eqref{zero} follows because $\omega$ and $\Omega$ are both positive, so the argument of the delta function in \eqref{c1} is never zero except when both vanish. 

For the massive case, we have:
\beq
C(\omega, \Omega)=\left(\frac{k}{2 \pi} e^{\frac{4 \pi}{\beta k}}\right)^{1-\Delta} e^{\frac{i \omega}{\pi}\left[\log \left(\frac{\beta k}{2 \pi}\right)-\gamma\right] e^{\frac{4 \pi}{\beta k}}} \delta\left(\frac{\omega k}{2 \pi}-\Omega\right)
\eeq

It follows that 
\begin{align}
  c_\o \alpha_{\omega \Omega} &= \left(\frac{k}{2 \pi} e^{\frac{4 \pi}{\beta k}}\right)^{1-\Delta} e^{\frac{i \omega}{\pi}\left[\log \left(\frac{\beta k}{2 \pi}\right)-\gamma\right] e^{\frac{4 \pi}{\beta k}}} \delta\left(\frac{\omega k}{2 \pi}-\Omega\right)\\
  \beta_{\omega \Omega} &=0
\end{align}

We find that for the evaporating black hole, $\beta_{\o \Omega}=0$ for both the massive and massless cases, which implies that the Hawking radiation vanishes. This is the expected result because the black hole itself will have disappeared in the late time limit.

\section{Summary}
In this paper, we have studied Hawking radiation in Jackiw-Teitelboim gravity minimally coupled to scalar fields. We considered both massive and massless scalars. 

We computed the spectrum for both a black hole at equilibrium as well as a black hole that has been attached to a bath. 

For a black hole in equilibrium we obtained a thermal spectrum, as expected. For a black hole out of equilibrium, we obtained the spectrum for two regimes --early time and late time. In the former case, we obtained a thermal spectrum in the semiclassical limit of the effective matter-gravity coupling $k \to 0$. This was the expected result as a black hole in the early stages of evaporation is close to an eternal black hole. We also computed corrections in the first order in $k$ for the massless scalar, and showed that the corrections result in deviation from thermal spectrum. 

In the late time limit, there are two cases. In the case where the bath has a non-zero temperature, we obtained once again a thermal spectrum, now at the temperature of the bath. This signifies the black hole settling to a new temperature. In this case, we did not restrict to the semiclassical regime and the computation takes into account all orders of $k$. We computed the corrections to the Bogoliubov coefficients away from the strict late-time limit and found no deviations from thermal spectrum.

Finally, we considered a black hole that evaporates completely. In this case, the Hawking radiation should vanish at late times as the black hole has disappeared. This was indeed found to be the case. 
\section{Acknowledgements}
We would like to thank Rinkesh Panigrahi and Shahbaz Dar for helpful discussions and suggestions.
NK is supported by SERB Start-up Research Grant SRG/2022/000970.

\bibliographystyle{unsrt}
\bibliography{main}

\begin{thebibliography}{10}

\bibitem{Callan:1992rs}
Curtis~G. Callan, Jr., Steven~B. Giddings, Jeffrey~A. Harvey, and Andrew Strominger.
\newblock {Evanescent black holes}.
\newblock {\em Phys. Rev. D}, 45(4):R1005, 1992.

\bibitem{Russo:1992ax}
Jorge~G. Russo, Leonard Susskind, and Larus Thorlacius.
\newblock {The Endpoint of Hawking radiation}.
\newblock {\em Phys. Rev. D}, 46:3444--3449, 1992.

\bibitem{deAlwis:1992hv}
S.~P. de~Alwis.
\newblock {Quantum black holes in two-dimensions}.
\newblock {\em Phys. Rev. D}, 46:5429--5438, 1992.

\bibitem{Giddings:1992ae}
Steven~B. Giddings and Andrew Strominger.
\newblock {Quantum theories of dilaton gravity}.
\newblock {\em Phys. Rev. D}, 47:2454--2460, 1993.

\bibitem{Jackiw:1984je}
R.~Jackiw.
\newblock {Lower Dimensional Gravity}.
\newblock {\em Nucl. Phys. B}, 252:343--356, 1985.

\bibitem{Teitelboim:1983ux}
C.~Teitelboim.
\newblock {Gravitation and Hamiltonian Structure in Two Space-Time Dimensions}.
\newblock {\em Phys. Lett. B}, 126:41--45, 1983.

\bibitem{Note1}
For earlier work on black hole thermodynamics in JT gravity see \cite {Lemos:1993qn,Lemos:1996bq}.

\bibitem{Almheiri:2019psf}
Ahmed Almheiri, Netta Engelhardt, Donald Marolf, and Henry Maxfield.
\newblock {The entropy of bulk quantum fields and the entanglement wedge of an evaporating black hole}.
\newblock {\em JHEP}, 12:063, 2019.

\bibitem{Penington:2019npb}
Geoffrey Penington.
\newblock {Entanglement Wedge Reconstruction and the Information Paradox}.
\newblock {\em JHEP}, 09:002, 2020.

\bibitem{Almheiri:2019yqk}
Ahmed Almheiri, Raghu Mahajan, and Juan Maldacena.
\newblock {Islands outside the horizon}.
\newblock 10 2019.

\bibitem{Geng:2020qvw}
Hao Geng and Andreas Karch.
\newblock {Massive islands}.
\newblock {\em JHEP}, 09:121, 2020.

\bibitem{Geng:2020fxl}
Hao Geng, Andreas Karch, Carlos Perez-Pardavila, Suvrat Raju, Lisa Randall, Marcos Riojas, and Sanjit Shashi.
\newblock {Information Transfer with a Gravitating Bath}.
\newblock {\em SciPost Phys.}, 10(5):103, 2021.

\bibitem{Geng:2021hlu}
Hao Geng, Andreas Karch, Carlos Perez-Pardavila, Suvrat Raju, Lisa Randall, Marcos Riojas, and Sanjit Shashi.
\newblock {Inconsistency of islands in theories with long-range gravity}.
\newblock {\em JHEP}, 01:182, 2022.

\bibitem{Geng:2023zhq}
Hao Geng.
\newblock {Graviton Mass and Entanglement Islands in Low Spacetime Dimensions}.
\newblock 12 2023.

\bibitem{Blommaert:2020yeo}
Andreas Blommaert, Thomas~G. Mertens, and Henri Verschelde.
\newblock {Unruh detectors and quantum chaos in JT gravity}.
\newblock {\em JHEP}, 03:086, 2021.

\bibitem{Giddings:1992ff}
Steven~B. Giddings and William~M. Nelson.
\newblock {Quantum emission from two-dimensional black holes}.
\newblock {\em Phys. Rev. D}, 46:2486--2496, 1992.

\bibitem{Hyun:1994na}
Seungjoon Hyun, Geon~Hyoung Lee, and Jae~Hyung Yee.
\newblock {Hawking radiation from (2+1)-dimensional black hole}.
\newblock {\em Phys. Lett. B}, 322:182--187, 1994.

\bibitem{Eyheralde:2022ayp}
Rodrigo Eyheralde.
\newblock {Hawking radiation from an evaporating black hole via Bogoliubov transformations}.
\newblock {\em Class. Quant. Grav.}, 39(22):225002, 2022.

\bibitem{Dey:2021vke}
Parijat Dey and Nirmalya Kajuri.
\newblock {Bulk reconstruction and Bogoliubov transformations in AdS$_{2}$}.
\newblock {\em JHEP}, 09:170, 2021.

\bibitem{Belin:2018juv}
Alexandre Belin, Nabil Iqbal, and Sagar~F. Lokhande.
\newblock {Bulk entanglement entropy in perturbative excited states}.
\newblock {\em SciPost Phys.}, 5(3):024, 2018.

\bibitem{Mertens:2022irh}
Thomas~G. Mertens and Gustavo~J. Turiaci.
\newblock {Solvable models of quantum black holes: a review on Jackiw\textendash{}Teitelboim gravity}.
\newblock {\em Living Rev. Rel.}, 26(1):4, 2023.

\bibitem{Maldacena:2016upp}
Juan Maldacena, Douglas Stanford, and Zhenbin Yang.
\newblock {Conformal symmetry and its breaking in two dimensional Nearly Anti-de-Sitter space}.
\newblock {\em PTEP}, 2016(12):12C104, 2016.

\bibitem{Engelsoy:2016xyb}
Julius Engels\"oy, Thomas~G. Mertens, and Herman Verlinde.
\newblock {An investigation of AdS$_{2}$ backreaction and holography}.
\newblock {\em JHEP}, 07:139, 2016.

\bibitem{Banks:1998dd}
Tom Banks, Michael~R. Douglas, Gary~T. Horowitz, and Emil~J. Martinec.
\newblock {AdS dynamics from conformal field theory}.
\newblock 8 1998.

\bibitem{Harlow:2011ke}
Daniel Harlow and Douglas Stanford.
\newblock {Operator Dictionaries and Wave Functions in AdS/CFT and dS/CFT}.
\newblock 4 2011.

\bibitem{Dobrev:1998md}
V.~K. Dobrev.
\newblock {Intertwining operator realization of the AdS / CFT correspondence}.
\newblock {\em Nucl. Phys. B}, 553:559--582, 1999.

\bibitem{Bena:1999jv}
Iosif Bena.
\newblock {On the construction of local fields in the bulk of AdS(5) and other spaces}.
\newblock {\em Phys. Rev. D}, 62:066007, 2000.

\bibitem{Hamilton:2005ju}
Alex Hamilton, Daniel~N. Kabat, Gilad Lifschytz, and David~A. Lowe.
\newblock {Local bulk operators in AdS/CFT: A Boundary view of horizons and locality}.
\newblock {\em Phys. Rev. D}, 73:086003, 2006.

\bibitem{Papadodimas:2012aq}
Kyriakos Papadodimas and Suvrat Raju.
\newblock {An Infalling Observer in AdS/CFT}.
\newblock {\em JHEP}, 10:212, 2013.

\bibitem{Kajuri:2020vxf}
Nirmalya Kajuri.
\newblock {Lectures on Bulk Reconstruction}.
\newblock {\em SciPost Phys. Lect. Notes}, 22:1, 2021.

\bibitem{Cadoni:1994uf}
Mariano Cadoni and Salvatore Mignemi.
\newblock {Nonsingular four-dimensional black holes and the Jackiw-Teitelboim theory}.
\newblock {\em Phys. Rev. D}, 51:4319--4329, 1995.

\bibitem{Hollowood:2020cou}
Timothy~J. Hollowood and S.~Prem Kumar.
\newblock {Islands and Page Curves for Evaporating Black Holes in JT Gravity}.
\newblock {\em JHEP}, 08:094, 2020.

\bibitem{Abramowitz}
M.~Abramowitz and I.~Stegun.
\newblock {\em {Handbook of Mathematical Functions with Formulas, Graphs, and Mathematical Tables}}.
\newblock Dover, New York, USA, 1964.

\bibitem{Bateman}
H.~Bateman and A.~Erdélyi.
\newblock {\em {Tables of Integral Transforms Vol.1}}.
\newblock McGraw Hill, New York, USA, 1954.

\bibitem{Lemos:1993qn}
Jose P.~S. Lemos and Paulo~M. Sa.
\newblock {Nonsingular constant curvature two-dimensional black hole}.
\newblock {\em Mod. Phys. Lett. A}, 9:771--774, 1994.

\bibitem{Lemos:1996bq}
Jose P.~S. Lemos.
\newblock {Thermodynamics of the two-dimensional black hole in the Teitelboim-Jackiw theory}.
\newblock {\em Phys. Rev. D}, 54:6206--6212, 1996.

\end{thebibliography}

\end{document}